\documentclass[conference]{IEEEtran}
\IEEEoverridecommandlockouts
\usepackage{cite}
\usepackage{amsmath,amssymb,amsfonts}
\usepackage{algorithmic}
\usepackage{graphicx}
\usepackage{textcomp}
\usepackage{xcolor}
\usepackage{enumitem}
\newcommand{\subscript}[2]{$#1 _ #2$}
\usepackage{xspace}
\usepackage{balance}
\usepackage{url}
\usepackage{listings}
\usepackage{multirow}
\newcommand{\smartparagraph}[1]{\noindent{\bf #1}\ }

\def\BibTeX{{\rm B\kern-.05em{\sc i\kern-.025em b}\kern-.08em
    T\kern-.1667em\lower.7ex\hbox{E}\kern-.125emX}}
\begin{document}

\newcommand{\Bindaas}{\emph{Bindaas}\xspace}

\title{Data Services with \Bindaas: \\RESTful Interfaces for Diverse Data Sources}

\author{\IEEEauthorblockN{Pradeeban Kathiravelu\IEEEauthorrefmark{1}$\quad$
Yusuf Nadir Saghar\IEEEauthorrefmark{2}$\quad$
Tushar Aggarwal\IEEEauthorrefmark{3}$\quad$ 
Ashish Sharma\IEEEauthorrefmark{1}}
\IEEEauthorblockA{\IEEEauthorrefmark{1}Emory University, Atlanta, GA, USA. \IEEEauthorrefmark{2}Vanderbilt University, Nashville, TN, USA}
\IEEEauthorblockA{ \IEEEauthorrefmark{3}BITS Pilani Hyderabad Campus, Hyderabad, India\\
\IEEEauthorrefmark{1}{\{firstname.lastname\}@emory.edu},
\IEEEauthorrefmark{2}nadirsaghar@gmail.com,
\IEEEauthorrefmark{3}f20150047@hyderabad.bits-pilani.ac.in}\\
\textbf{Pre-print Submitted to the IEEE BigData 2019 Conference}}

\maketitle
\begin{abstract}

The diversity of data management systems affords developers the luxury of building systems with heterogeneous systems that address needs that are unique to the data. It allows one to mix-n-match systems that can store, query, update, and process data, based on specific use cases. However, this heterogeneity brings with it the burden of developing custom interfaces for each data management system. Developers are required to build high-performance APIs for data access while adopting best-practices governing security, data privacy, and access control. These include user authentication, data authorization, role-based access control, and audit mechanisms to avoid compromising the security standards mandated by data providers. 

In this paper, we present \Bindaas, a secure, extensible big data middleware that offers uniform access to diverse data sources. By providing a standard RESTful web service interface to the data sources, \Bindaas exposes query, update, store, and delete functionality of the data sources as data service APIs, while providing turn-key support for standard operations involving security, access control, and audit-trails. \Bindaas consists of optional features, such as query and response modifiers as well as plugins that implement composable and reusable data operations on the data. The research community has deployed \Bindaas in various production environments in healthcare. Our evaluations highlight the efficiency of \Bindaas in serving concurrent requests to data source instances. We further observe that the overheads caused by \Bindaas on the data sources are negligible.

\end{abstract}

\section{Introduction}
\label{sec:intro}

The increasing volume and variety of big data have steered innovation in databases community. Several NoSQL and relational databases are purpose-built for various purposes of the big data applications~\cite{ntarmos2014rank}. The diversity in data sources in terms of data storage, access, and processing has helped enhance the performance of the applications in various domains. Databases continue to become more efficient and scalable, with guarantees of ACID (atomicity, consistency, isolation, durability) properties~\cite{vogels2009eventually}. Furthermore, several distributed data stores are developed, considering consistency, availability, and partition-tolerance aspects~\cite{han2011survey}. Hence, the increasing number and diversity of relational databases, as well as the NoSQL data sources, is indeed a promising trend.

\smartparagraph{Challenges of Data Source Diversity:} Nevertheless, such diversity comes with several challenges~\cite{chen2013big}. The data source interfaces do not follow a universal standard, each having their structures and query languages. Therefore, the data sources lack a standard interface or a communication channel to enable access and processing of data uniformly stored in them. This state of affairs imposes several challenges and overheads on the big data application developers and users. First, the developers and users should be knowledgeable of the interfaces of multiple data sources and applications. Second, the lack of a common interface makes big data applications that consume data from data sources to be tightly coupled to the underlying backend data sources. A big data application that communicates with data sources directly through their diverse interfaces becomes increasingly hard to maintain, more so when the application needs to consume data from different data sources. Third, adopting an existing application built for a specific data source for another data source is a complex engineering undertaking. Therefore, such an implementation limits its reusability across multiple data sources. A middleware framework that offers a common interface to the diverse data sources will mitigate these challenges, while not restricting the innovation of the data sources and the big data applications.

\smartparagraph{Limitations in Big Data Federation Middleware:} A framework that provides interoperable interfaces, with minimal overhead in terms of performance and operational cost to the big data application developers, is still lacking in practice. Web services offer standard interfaces, typically for smaller, computation-intensive applications. \textit{Data services}~\cite{li2004event} aim to bring web service interfaces to the data sources, to provide them with service-based access in a wide area network. Nevertheless, expecting the application developer to build service interfaces for each of the data sources where the data is hosted, incurs an overhead of repeated development effort. Big data access and federation middleware platforms eliminate the overhead and repeated attempts of developing such interfaces from application developers. However, such existing platforms limit their scope to specific data sources, rather than providing a generic service-based interface to data sources of several vendors. They also have massive resource requirements or complex configurations. Ironically, they also expect users to learn their interface syntax and semantics. Such middleware frameworks also cause overheads in converting data formats from the data sources.

\smartparagraph{Motivation:} Given the above premises, we aim to address the following research questions in this paper: 

\begin{enumerate}[label=(\subscript{RQ}{{\arabic*}})]
\setlength{\itemindent}{1.8em}
\item Can we expose the data stored in diverse data sources as data service APIs with negligible overhead by building a web services middleware?

\item Can such middleware support concurrent requests to multiple data sources without causing bottlenecks?

\item Can we provide secured access to the data sources to retrieve data and read existing data from the data sources, by using the access control mechanisms provided by the data sources themselves as well as third-party authentication providers?

\item Can we write chains of functions via a common interface to alter the database queries before their execution on the data sources as well as the response from the data sources for the query, without repeatedly developing data source operations for the same functionality?

\item Can such datasource configurations be shared and used among researchers beyond organization boundaries to minimize repeated configuration and development efforts?
\end{enumerate}

\smartparagraph{Contributions:} This paper aims to answer the identified research questions. The main contributions of this paper are:

\begin{enumerate}
        \setlength{\itemindent}{1.8em}
\item A data service approach to diverse relational, NoSQL, and other data sources, by offering standard web services interface to the data sources. ($RQ_1$)
\item An efficient modular architecture to enable secure, extensible big data middleware.  ($RQ_2$ and $RQ_3$)
\item Reusable chains of data operations as modifiers on the queries and query results as well as optional plugins. ($RQ_4$ and $RQ_5$)
\end{enumerate}

\Bindaas is an extensible open-source middleware that provides unified access and processing for various data sources via ready-to-use RESTful web service~\cite{richardson2008restful} interfaces. Supported database systems include MongoDB, MySQL, Postgres, and IBM DB2. In addition to such classic databases, \Bindaas also offers RESTful interfaces to other data storage systems such as web-based data storage via its HTTP-based access. Interfaces and abstract classes are also available for developers who wish to build such interfaces for more data sources with minimal effort. For example, its abstract interface for SQL has been extended for big data frameworks such as Apache Drill~\cite{hausenblas2013apache}, Google BigQuery~\cite{tigani2014google}, and Apache Hive~\cite{thusoo2009hive}.

Users can use \Bindaas to create RESTful interfaces, in a declarative manner, and provide access to data sources, and let users query, update, store, and delete data, securely. \Bindaas adopts a modular architecture, thus allowing the development and deployment of custom user-defined features. Our evaluations highlight that it causes minimal overhead during data access and processing, compared to accessing and processing the data directly from the data sources. It can support a large number of concurrent service requests to the backend data sources, without dropping the requests.

\smartparagraph{\textbf{Paper organization:}} In the following sections, we further discuss the \Bindaas middleware. Section~\ref{sec:background} discusses the state of the art and related work that proposes interoperable interfaces for diverse data sources. Section~\ref{sec:arch} elaborates the \Bindaas architecture in detail, discussing the \Bindaas workflow and security mechanisms. Section~\ref{sec:impl} presents the implementation details of \Bindaas. Section~\ref{sec:eval} discusses our evaluations on the \Bindaas middleware. Finally, Section~\ref{sec:concl} concludes the paper with a summary of our findings and future work.
\section{Background and Related Work}
\label{sec:background}

Research efforts on improving the interoperability of diverse data source interfaces have been promising. However, state of the art does not provide an extensible framework that offers standardized RESTful interfaces to the diverse data sources with minimal management and performance overheads. In this section, we will look into a few research works that are related to the \Bindaas approach.

\smartparagraph{Data Federation Middleware:} There have been middleware frameworks aiming to provide communication and federation of diverse data sources. Enterprise Service Bus (ESB)~\cite{chappell2004enterprise} refers to a class of middleware frameworks that provide a communication mechanism and interoperability to different data sources and systems. ESBs tend to be large and heavy in configurations, loaded with functionalities that are unessential for big data frameworks. WSO2 builds an ESB~\cite{indrasiri2016introduction} and a Data Services Server as OSGi platforms. The WSO2 architecture lets the users integrate and build only the minimum functionalities necessary. WSO2 Data Services Server exposes data sources with composable SOA interfaces. The modular architecture of WSO2 Data Services Server enables it to be light-weight with minimal performance overheads and resource requirements~\cite{hamlen2013data}.

Unified access to legacy and non-traditional data sources to coordinate and operate with the current enterprise data sources is a long-researched research topic. Legacy data source wrappers~\cite{roth1997don} have been proposed to enable integrating legacy data sources into contemporary applications, rather than altering the applications to work with the legacy monolith data stores. OGSA-DAI~\cite{antonioletti2005design} supports access and integration capabilities across multiple relational databases, files, XML data stores, and web services. OGSA-DAI aims to reduce the developer overhead on identifying data location, structure, and communication mechanisms and focus on application-specific constructs such as data processing and analysis instead. 

\smartparagraph{Optimizing Database Queries:} There have been several efforts on optimizing database queries across diverse database systems. MOCHA~\cite{rodriguez2000mocha} is a query processing middleware framework that integrates various data sources over the network, optimizing the queries. Garlic~\cite{haas1997optimizing, carey1995towards} focuses on optimizing query performance across diverse data sources through a common interface. While these related works also have focused on providing unified access to diverse data sources, they are often difficult to extend. Moreover, they do not consider the practical challenges concerning access control, security, and extensibility. \Bindaas limits its scope to providing interoperable data source query execution with little to no overheads. Optimizing database queries is out of the scope of \Bindaas.

\smartparagraph{Unified Data Access Interfaces:} 
Middleware frameworks aiming to enhance the interoperability of the data sources have also been developed specifically for certain application domains or data source vendors. Some middleware platforms limit their focus to offer web services or RESTful interface to just one or a few data sources~\cite{roe2010restful} -- for example, interoperable interfaces spanning IBM data sources~\cite{carey1998data}. They do not provide capabilities for such interfaces to be extended to any data source, beyond a specific vendor.

Application-specific data access and integration middleware platforms tend to be difficult to generalize as general-purpose middleware platforms, due to their tight-coupling with the data sources and their assumptions concerning the data. Data-Source Interoperability Service (DSIS)~\cite{pang2015data} wraps data sources, including traditional relational and NoSQL databases, decision support systems, and Auto-ID devices into a Service-Oriented Architecture with web services engines and web services registry. DSIS, however, limits its focus to Enterprise Information Systems (EIS). miRMaid~\cite{jacobsen2010mirmaid} exposes diverse data sources of microRNAs as read-only RESTful resources. The modular architecture of miRMaid can indeed facilitate extending it for various application domains with several other data sources. However, its approach to data as read-only resources rather than services that can read and write to the data sources limits its applicability. RESTful APIs have been developed for the Bacterial Isolate Genome Sequence Database (BIGSdb), with query, update, store, and delete data entries with proper authentications~\cite{jolley2017restful}. While this approach is similar to \Bindaas, we note that the design of \Bindaas makes it extensible while also supporting specifics of each application domains incorporated into \Bindaas instances without additional development effort. \Bindaas aims to address the shortcomings in the current big data access middleware, as an extensible modular big data middleware.
\section{Solution Architecture}
\label{sec:arch}

\Bindaas is an extensible, modular, middleware platform that allows users to create RESTful interfaces for their data. A user can connect their data sources with a running \Bindaas instance with minimal configuration. It's architecture aims to provide secure access to various data sources with minimal downtime and overhead. \Bindaas exposes its functionalities of creating APIs for data sources as well as various administrative functions through a web console and a RESTful interface.
\begin{figure}[ht]
    \begin{center}
            \resizebox{0.9\columnwidth}{!}{
                \includegraphics[width=0.9\textwidth]{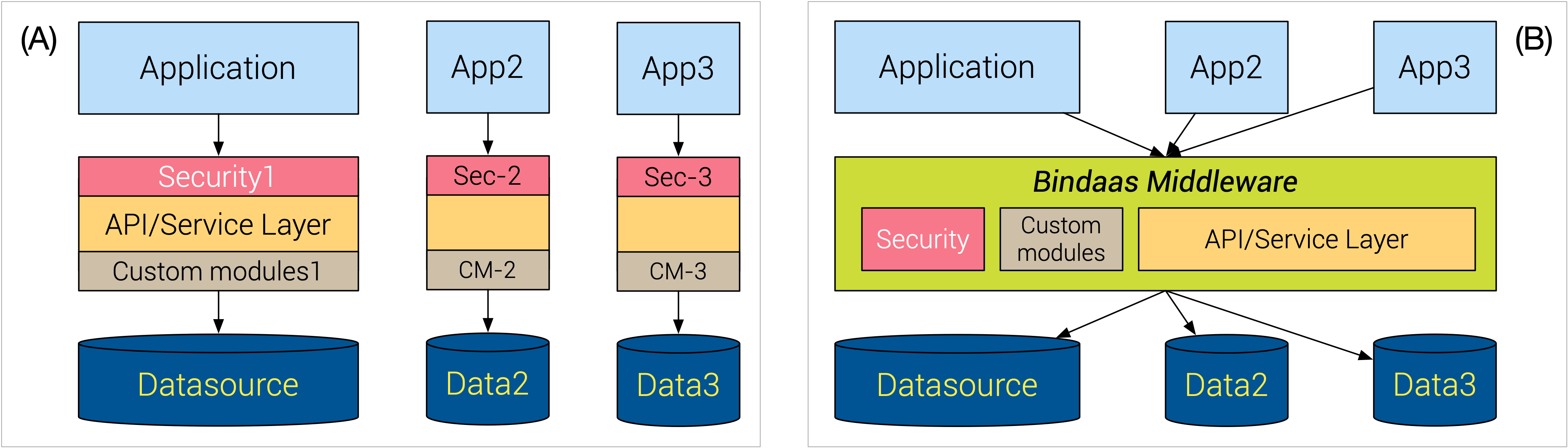}
            }
        \end{center}
        \caption{(A) Classical data applications require repeat implementations for API, security etc. (B)\Bindaas simplifies and harmonizes the API development process through shared best practices for common operations}
        \label{fig:both}
    \end{figure}
    
\subsection{\Bindaas Approach} 
\Bindaas middleware provides standard RESTful interfaces to diverse data sources. It thus reduces custom-code tailor-made for multiple data sources by the application developers. Figure~\ref{fig:both} summarizes how \Bindaas big data middleware avoids the need for such glue-code or custom modules developed by different application developers to enable communication between the big data application and the data sources. It does not limit the big data applications to a specific data source. Hence, \Bindaas aims to enable seamless migration of big data applications to use different data sources.
\Bindaas supports multiple datasource providers. A datasource provider is an implementation that allows the middleware to connect to a specific database engine. Some datasource providers are specific to a given database engine, whereas a few others are generic and abstract and hence are extended by multiple other datasource providers. For example, the \Bindaas MongoDB datasource provider offers a RESTful interface to MongoDB, whereas the \Bindaas HTTP datasource provider provides an interface to web-based data sources. On the other hand, a few other \Bindaas datasource providers offer a more generic interface to multiple data sources. Such generic interfaces are extended by other datasource providers for the concrete implementation for a specific data source. For example, MySQL, Postgres, IBM Db2, and Drill datasource providers offer RESTful interfaces respectively to MySQL, PostGres, and IBM Db2 databases as well as Apache Drill. These datasource providers extend the Generic SQL datasource provider. Apache Drill itself can be configured to query multiple data sources. \Bindaas also provides code stubs, abstract classes, and frameworks to enable quick implementation of more datasource providers by the community, to offer RESTful interfaces to other data sources.

A running instance of \Bindaas can be configured to connect with multiple instances of various data sources. Each configured data source creates a project file. These project files can be created via the web user interface, manually, or programmatically via the \Bindaas administrative APIs. The OSGi-based hot deployment enables \Bindaas to reflect the changes in the projects without restarting the middleware. Users can create, maintain, and use multiple project files in \Bindaas. Furthermore, once a project file is created, it can be shared with other users. The possibility to share and reuse the project files across multiple \Bindaas instances avoids the need to configure the data source connectivity numerous times.

\Bindaas consists of a set of users conceptually belonging to one or more categories as depicted by Table~\ref{table:both}. The \Bindaas administrator configures \Bindaas and executes it. The API developer creates project files for her data sources. The API user invokes the APIs created by the API developers to consume the data sources. In small-scale deployments of \Bindaas, these user roles/categories overlap. For example, the \Bindaas administrator may also be the only one who creates the project files. Similarly, the API user can be the same person as the API developer, if the APIs are developed and consumed by the same user.

\begin{table}[!t]
\centering
\caption{The \Bindaas Users}
\label{table:both}  
\begin{tabular}{|l|l|l|}
\hline
\multicolumn{2}{|c|}{\textbf{\Bindaas User}}            & \textbf{Description}                                                                                                  \\ \hline \hline
\multicolumn{2}{|l|}{\Bindaas Administrator}           & \begin{tabular}[c]{@{}l@{}}Configures and maintains an \\     instance of \Bindaas middleware.\end{tabular}    \\ \hline
\multirow{2}{*}{\Bindaas User} & API Developer & \begin{tabular}[c]{@{}l@{}}Creates \Bindaas APIs for \\       her data sources.\end{tabular}                   \\ \cline{2-3} 
                              & API User      & \begin{tabular}[c]{@{}l@{}}Consumes the \Bindaas APIs to  \\ access and process the data sources.\end{tabular} \\ \hline
\end{tabular}
\end{table}

\subsection{\Bindaas Workflow}

\begin{figure}[ht]
    \begin{center}
            \resizebox{0.8\columnwidth}{!}{
                \includegraphics[width=0.8\textwidth]{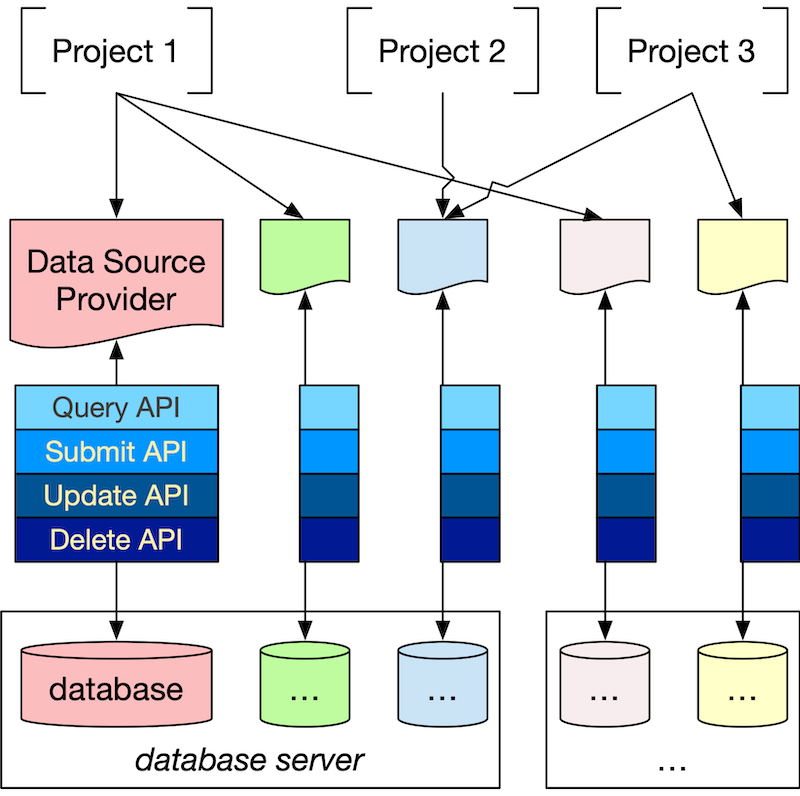}
            }
        \end{center}
        \caption{The \Bindaas Project Structure. Each project can map to multiple databases (data providers) and specific CRUD APIs, for these databases, are stored in projects}
        \label{fig:bworkflow}
    \end{figure}

In \Bindaas, an API for a data source is organized as Projects. A project consists of one or more data providers and a single \Bindaas instance can include multiple projects. Data providers in a project are bound to a specific database. As the name suggests, a data provider describes the source of the data you like to interact with. In most cases the data resides in a database (relational or NoSQL). Bindaas needs to know where to get the data from. Depending on the database system Bindaas requires some information such as: host, port, credentials, other miscellaneous information to establish a connection.  Once a connection is establish a Data Provider gets created. This entire  information is stored, for each project, in a project file. The API users can then interact with and query the data sources via the defined \Bindaas data source APIs. Figure~\ref{fig:bworkflow} elaborates this workflow of creating and consuming projects in \Bindaas.

Each datasource provider in \Bindaas consists of a query endpoint, update endpoint, delete an endpoint, and submit endpoint. These endpoints query the data source, update an existing entry in the data source, delete existing data from the data source, and create new data entries in a data source to store data, respectively. These endpoints follow the CRUD (Create, Retrieve, Update, and Delete) patterns of the REST architecture. An HTTP POST action invokes the Submit API. Similarly, HTTP GET, PUT, and DELETE actions, respectively invoke the Query API, Update API, and Delete APIs.

\smartparagraph{Query Templates:}\Bindaas aims to offer the API users (, i.e., the users who need to access the data sources for their big data applications) a unified RESTful interface to the data sources. The query API is used to query the data in the database linked to the data provider. It describes the execution query, together with the user-provided parameters in the query. It also indicates the output format of the outcomes, such as JSON, CSV, and HTTP. API developers have the optional ability to assign metadata for each API they create. This metadata can store information about the API, including query parameters as well as information about the structure of the result set. The query API receives the additional details required by the data source, at the time of the project creation, or when the project files are later updated. \Bindaas middleware is agnostic to any query language. However, the API developer writes a \Bindaas query template in the language supported by default by the specific datasource provider. For example, when working against a data provider that connects to a relational database, the query is written in SQL. Similarly, when working with MongoDB, the query is written in JSON. This query template is then parsed by the middleware and used to generate a RESTful API for an end-user, regardless of the underlying data source. Therefore, the API users can access the data sources without using the query language specific to the data source.

\Bindaas supports all the queries of Mongo and SQL data sources. Internally, it passes on the queries as they are, with no modifications to the queries, using the Java interface of the data sources. The outputs that are resultants of the endpoint executions are handled by a relevant `Handler,' before returning the outcome to the user. \Bindaas consists of a Submit Handler, Query Handler, Update Handler, and Delete Handler for each datasource provider. They respectively manage the create (POST), retrieve (GET), update (PUT), and delete (DELETE) actions.

\smartparagraph{Bind Variables:} The API developers should define and describe the runtime parameters of each query, to enable the user to execute the queries with different values for the query parameters. We call these parameters, ``Bind Variables''. A bind variable is annotated with the dollar sign (`\$'). For example, `patientID' is a bind variable in the below \Bindaas query for MySQL datasource provider.

{\fontsize{8}{8}\selectfont                  
\begin{lstlisting}
SELECT * FROM PATIENT_TABLE 
WHERE PATIENT_ID = $patientID$
\end{lstlisting}
}

\Bindaas seeks and matches this regEx. It then extracts and replaces it by the value supplied by the API user. For example, the query: http://example.org?[query\_param\_list] is executed using the following RESTful Call:

{\fontsize{8}{8}\selectfont                  
\begin{lstlisting}
HTTP GET http://server:port/services/{project}/
    {data_provider}/query/{query_name}?
    {query_param_list}
\end{lstlisting}
}

\Bindaas is a framework that can be extended with minimal configuration effort and without boilerplate and glue code. \Bindaas consists of a modular architecture, based on the OSGi~\cite{alliance2003osgi} framework. Each feature is built as an OSGi bundle and deployed in the OSGi execution environment of \Bindaas. Additional features such as throttling and rate-limiting can be developed by the users and deployed dynamically into a running \Bindaas instance. \Bindaas consists of several plugins and modifiers, in addition to the datasource providers. The plugins are optional functionality such as rate limiting and load balancing that can execute on the data that is retrieved from the data sources. In addition to writing more datasource providers to integrate more data sources, the users can also write their custom plugins and modifiers. The modifiers can be a Query Modifier, Query Result Modifier, or a Submit Payload Modifier. The modifiers alter the user query or the response returned by the data source for the query. The API developer can define modifiers on her workflow - from the existing \Bindaas modifiers or she can develop her modifiers.

The \textbf{Query Modifier} alters the user query before the query is executed on the data. When a user invokes a \Bindaas RESTful query interface defined on the data source, any query modifiers defined by the API developer executes on the query. For example, a query modifier, for a relational database, could be used to inspect a query for possible injection. Another modifier could be used to perform access control based on the query and the API user who invoked the API. The query can be changed or augmented based on the authorization granted to the API user who invoked the API, using the query modifier.\par
The \textbf{Query Result Modifier}, on the other hand, alters the results of a query. Blob Download Plugin, Image Download Query Result Modifier, and Generic Output Chainer Query Result Modifier are a few notable Query Result Modifier implementations. The Blob Download Plugin enables binary output of a query to be saved in the desired formats. The Image Download Query Result Modifier processes raw JSON produced by the Query Handler into corresponding images, then finds the respective file paths of the images, and produces a zip archive consisting of the images. Queries can also be chained to send the outcome of a previous query as an input to the next query, thus avoiding the need to persist the intermediate outcomes in a big data workflow. The Generic Output Chainer Query Result Modifier implements this chaining capability to make a workflow. Query result modifiers can also perform analytics on the results returned by the query. A query can further have chains of query modifiers preceding the query execution, and followed by a series of query results modifiers.\par
\begin{figure}[ht]
    \begin{center}
            \resizebox{0.95\columnwidth}{!}{
                \includegraphics[width=0.95\textwidth]{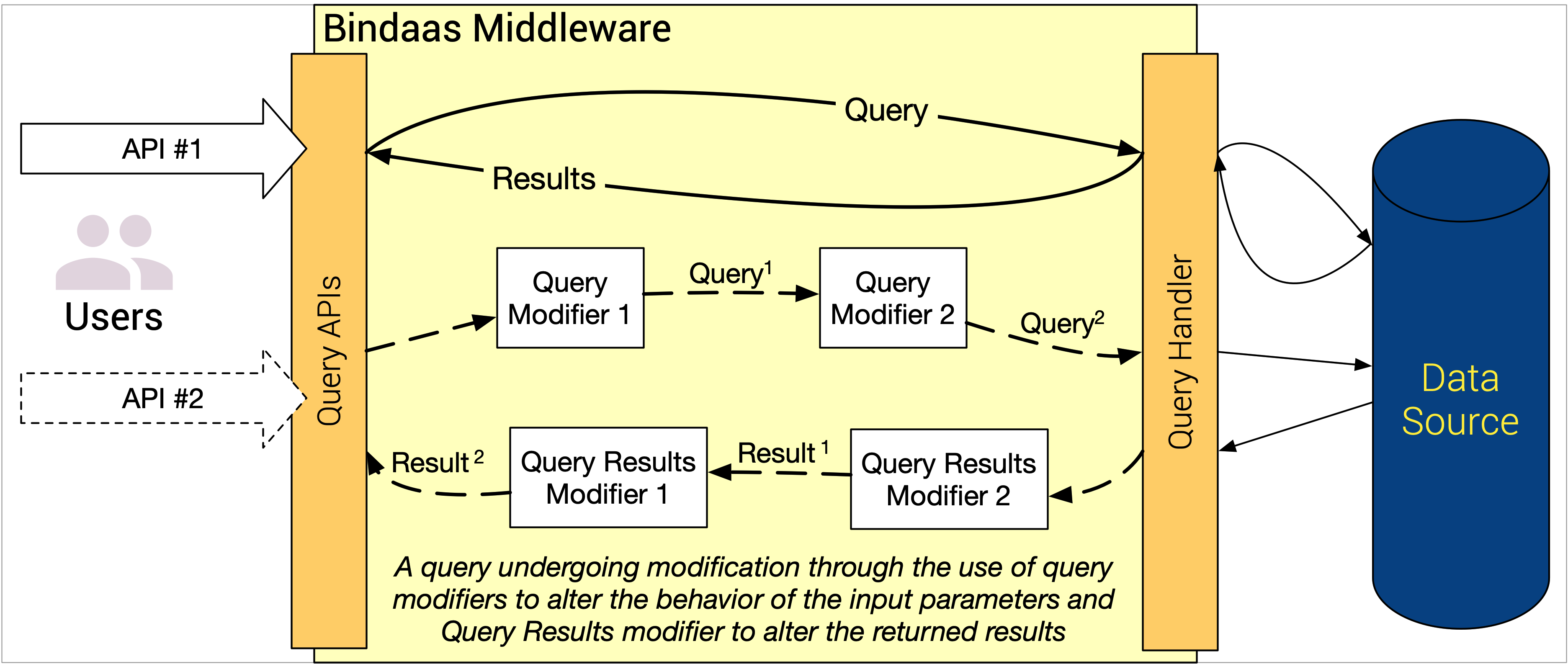}
            }
        \end{center}
        \caption{The use of modifers to alter the behavior of query and query results}
        \label{fig:flow}
    \end{figure}
The \textbf{Submit Payload Modifier} similarly the payload of a submit API execution workflow before returning the payload that was the result of an HTTP POST query to the Submit Handler and eventually to the user. For example, in the cancer imaging workspace images require anonymization and data cleaning procedures before they are submitted. Such procedures can be incorporated as the Submit Payload Modifiers. Figure~\ref{fig:flow} highlights the data flow of \Bindaas with the modifiers. The dotted lines indicate that the query modifiers and query results modifiers are components of API\#2. Each query modifier or a query results modifier can have one or more instances, each providing different functionality, in the data flow. The workflow of the POST requests is similar to the one depicted, with Submit Handler and Submit Payload Modifier, in place of Query Handler and Query Modifier respectively. Query Results Modifier is common to all types of operations on the data sources.\par
\smartparagraph{Persisting Data:} \Bindaas stores its persistent data in an SQLite data store. Authentication and access details, including API key, are persisted in the data store by default. The project files are stored directly in the file system to easily manage and update them without having to use the database queries. The audit trails consisting of access logs are stored in the data store in addition to the log files, to enable efficiently executing complex queries on them. The persistent data storage interface can be implemented to use other data sources as well. For example, if \Bindaas has to persist a much larger internal data, such as authentication details, the interface can be extended to use MongoDB as the persistent data store. Similarly, an In-Memory Data Grids~\cite{gaur2016system} such as Hazelcast or Infinispan, can replace the SQLite data store in favor of a better performance.

\subsection{Authentication, Authorization, and Access Control}
A middleware that provides access to data sources should ensure the data security is not compromised. \Bindaas aims to exploit the authentication mechanism provided by the data sources in accessing data, while also incorporating the industrial security best practices. \Bindaas consists of Security Providers, each implementing various security features such as LDAP integration.

\begin{figure}[ht]
    \begin{center}
            \resizebox{0.9\columnwidth}{!}{
                \includegraphics[width=0.9\textwidth]{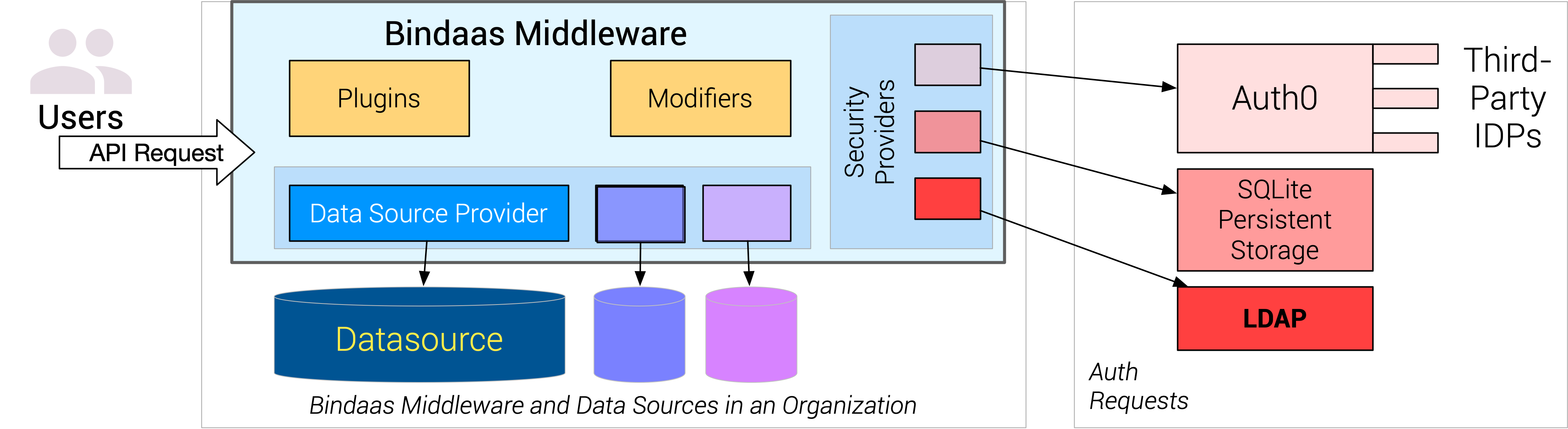}
            }
        \end{center}
        \caption{\Bindaas Deployment}
        \label{fig:depl}
    \end{figure}

\Bindaas offers several alternative authentication mechanisms which can be enabled by the administrator by using the \Bindaas configuration file. The default authentication option is to configure \Bindaas with API key-based authentication. \Bindaas also offers JWT (JSON Web Token) based authentication~\cite{jones2013json}. \Bindaas can easily be extended to incorporate custom authentication approaches as well. The administrators can also disable \Bindaas features such as authentication and rate-limiting and use an API gateway to provide such functionality to the user queries before the queries reach the \Bindaas middleware. \Bindaas preserves the additional HTTP headers in the service requests. Therefore, the headers can be used to selectively perform additional operations using modifiers later in the data flow. External API gateways and load balancers can be used to have replicated and load balanced \Bindaas deployment, to enhance further the \Bindaas performance in managing numerous concurrent queries.  The service invocations must hold the adequate authentication headers unless the authentication is disabled in the \Bindaas server through its configuration options available to the administrator. Figure~\ref{fig:depl} elaborates a typical \Bindaas deployment with the data sources deployed in the organization and the third-party authentication providers. 

\Bindaas uses Auth0 Identity Infrastructure as a Service framework~\cite{pace2016identity} for its JWT based authentication. \Bindaas can be configured to generate JWTs for the users from its middleware or validate JWTs generated by third-party authentication sources. Interfacing with Auth0 ensures that \Bindaas can seamlessly authenticate via any third-party authentication provider rather than being limited to a specific one. Thus, \Bindaas interconnects with third-party OpenID Connect~\cite{sakimura2014openid} services such as Google seamlessly. 

\Bindaas generates the API keys when a user logs in to the \Bindaas web console or through by invoking its RESTful API for the API key generation by the administrator. \Bindaas maintains API keys for its users locally, which can be viewed from the \Bindaas client. The API keys are issued with an expiry date, until which the service executions are permitted by the user that holds the API keys. The \Bindaas Users can fetch their JWT tokens or API keys by accessing the web console of \Bindaas. The administrators can view all the users, their access tokens, and their defined APIs, whereas the users can view only their own.

\Bindaas holds the authorization details in a data store, file system, or an LDAP server~\cite{howes2003understanding}. 
By default, \Bindaas holds the authentication details including the roles of the users in either an SQLite filesystem-based data store or an LDAP directory. When a service request comes with the authentication header, the requests are validated against this stored user data in \Bindaas. The authentication decision is cached for a certain period, as configured by the \Bindaas administrator. Caching of the authentication decisions aims to minimize performance overheads.

\Bindaas also provides extensible access control to data sources. The datasource providers are equipped with extension points that support further fine-grain access control to the data sources, rather than limiting them to the levels of tables or collections. For example, an API user can be given access to only certain `project' entries. Here `project' is an attribute, and only the specific values of projects are accessible for certain users. A complete implementation of such finer access control is difficult to implement. Therefore, we built a reference implementation for TCIA projects. In the \Bindaas reference implementation for TCIA, only certain projects/collections are accessible for the public while limiting access to other projects in the table to a subset of the authenticated users.

Databases can be configured to limit concurrent queries from a user or an application. However, that would require extensive tweaking to the database server, which ironically would lead to performance degradations. \Bindaas facilitates rate limiting and throttling the database connections and concurrent requests from the users, based on their roles in the LDAP directory in \Bindaas deployment together with the information embedded in their JWT header information.

\subsection{\Bindaas Production Deployments} 
\Bindaas is deployed in production to manage several enterprise data sources for multiple application scenarios. 
The Cancer Imaging Archive (TCIA)~\cite{clark2013cancer} is NCI’s primary resource for distributing images and related data to support Cancer Research. TCIA is visited each month by more than 4000 users from around the world, actively supports 8000 registered users and many more anonymous users. \Bindaas is the middleware that drives the TCIA ecosystem. The underlying metadata is stored in relational databases with an underlying object store. The APIs allow users to query and retrieve large volumes of data in a highly efficient manner. The API is frequently used by researchers for bulk download and replication, as well as to build custom SDKs in Python and R Bioconductor packages. The various cancer images are stored as DICOM (Digital Imaging and Communication in Medicine)~\cite{mildenberger2002introduction} objects, in an object store. Consequently, a \Bindaas result modifier plugin is used to stream data off the storage, and onto an end-user via Java zipstreams.

A primary driver behind the development of \Bindaas was a need to standardize the security and auditing mechanisms. These are crucial components of any web service that is providing access to healthcare data. \Bindaas has been used to develop a platform for real-time prediction of the onset of sepsis~\cite{shashikumar2019deepaise}. In this platform, live data is acquired from electronic medical records, and streamed to a deep learning algorithm, that examines a host of features and predicts the liklihood of a patient developing sepsis within the next 8 hours. The data is stored as time-series documents and updated to include the computed probability, and stored in MongoDB for a clinican facing dashboard. \Bindaas is used to develop the various microservices, and is responsible for providing access to the various services, as well as managing security, access control, audit trails, and HTTPD~\cite{luotonen1995w3c}-style access logs. 

\Bindaas is responsible for supporting the core tech stack that powers the caMicroscope~\cite{saltz2017containerized} platform for pathomics and digital pathology. This includes metadata management of digital pathology images, as well as the management of various pathomic features. Of note is the size and scale of pathomics. A single whole slide digital pathology image is 2-4GB in size, and typical datasets can consist of a few thousand images. Pathomic features, from each image, consist of hundreds of thousands of segmented objects and associated features. In caMicroscope, all this data is managed in large deployments of MongoDB, and accessed via APIs that are powered by \Bindaas. When a user is exploring an image, with overlays of associated pathomic features, the API layer is processing and returning tens of thousands of JSON documents without any loss in interactivity. caMicroscope and the underlying API ecosystem, powered by \Bindaas, is responsible for various projects including a virtual tissue repository for NCI  Surveillance, Epidemiology, and End Results (SEER) Program. These are production grade systems that are used in ongoing studies and such production deployments highlight the stability, performance, and functionality of \Bindaas.

\section{Implementation}
\label{sec:impl}
We developed \Bindaas as an open-source big data middleware with Java 1.8\footnote{The source code, binaries, and Docker images of \Bindaas can be found at \url{https://github.com/sharmalab/bindaas/} with recent developments in appropriate branches.}. \Bindaas uses Eclipse Equinox~\cite{mcaffer2010osgi} as its core OSGi framework. It uses Apache Maven 3~\cite{miller2010apache} to manage and build its modules, and Apache CXF~\cite{balani2009apache} as its core web services middleware. It leverages the Spring framework~\cite{johnson2004spring} as its Inversion of Control (IoC) container to manage its application beans and their dependencies. \Bindaas uses Eclipse Gemini Blueprint to enable dynamic updates to existing modules, as well as adding and removing the OSGi bundles dynamically. The OSGi framework of \Bindaas enables multiple versions of a module to coexist simultaneously at run time. Apache Felix Gogo provides an interactive OSGi shell.

\Bindaas also consists of optional functionalities in addition to offering interfaces to the data sources. Notable examples include: i) the email notifier service that alerts the administrator for specific actions predefined by the administrator, such as a new user signing up, and ii) role-base access control to database tables/collections in a finer-grain at the entry/document level. \Bindaas logs its events extensively, using Log4j2 with several levels of priority. It stores the logs in rolling log files as well as in an SQLite filesystem-based database. The logs include \Bindaas middleware logs, logs from the underlying database servers, and the access logs indicating the API users who accessed the \Bindaas APIs. 

\subsection{Configuring \Bindaas}

Administrators can configure \Bindaas quickly based on their requirements, using various approaches. The administrator can configure \Bindaas by using the web console, via its configuration files of JSON format at the time of \Bindaas startup, by invoking the RESTful APIs of \Bindaas, or through client applications that consume the \Bindaas APIs. The administrators can also disable the web console and manage \Bindaas entirely via the other three approaches.

\Bindaas defines project files, consisting of configuration details of the data source and the query information. Each project file defines one or more datasource providers, for each data source server that it intends to interface with. Each datasource provider in the project file has one or more endpoints, referring to CRUD on the data sources. The project file follows a JSON format, regardless of the underlying data store. A sample project file with minimal details is shown.

{\fontsize{8}{8}\selectfont                  
\begin{lstlisting}
{ "profiles": {
    "Provider1": {
      "dataSource": {
        "host": "127.0.0.1",
        "port": "27017",
        "db": "EmployeeDB",
        "collection": "Employee",
        "authenticationDb": "admin",
        "username": "",
        "password": "",
        "initialize": false  },
      "queryEndpoints": {      },
      "deleteEndpoints": {      },
      "submitEndpoints": {
        "UploadEmployeeDetails": {
          "type": "FORM_DATA",
          "properties": {
            "inputType": "CSV",
            "csvHeader": [""]  },
          "name": "UploadEmployeeDetails",
        }  },  "providerId": "MongoDBProvider",  }  }, }\end{lstlisting}
}

The exact properties to define in the project file for a datasource provider, including the mandatory and optional features, depend on the respective database server that the project connects with. The authentication details can be embedded in the project files or retrieved at run time from the queries. 

\subsection{Security}
We implemented \Bindaas with a wide range of authentication and authorization alternatives. We configured OpenDJ~\cite{varanasi2013java} light-weight Java LDAP directory for several of our \Bindaas production deployments. \Bindaas does not have any tight-coupling or assumptions on the use of a specific LDAP server. The users can configure \Bindaas to use a specific LDAP directory (if any), through the \Bindaas configuration file in the JSON format.

\smartparagraph{Configuring with Auth0:} \Bindaas leverages Auth0 for seamless interfacing with various authentication providers. We developed \Bindaas binaries and containers with the configurations for Auth0. To configure \Bindaas to authenticate with Auth0, first, the \Bindaas administrator should create a new account in auth0, and register a new tenant in Auth0. Then, she should create a new application and configure it for the Authorization Code Flow~\cite{hardt2012rfc}. The Authorization Code Flow~\cite{hardt2012rfc} returns an access token to use with the custom endpoint we configured above. Finally, she should create a new endpoint for \Bindaas, configuring connections for multiple identity providers. We also developed custom javascript to modify the access token to incorporate roles into them to allow role-based access to \Bindaas data service APIs and data. \Bindaas \textit{DefaultJWTManager} verifies these tokens using RSA256 whenever any service endpoint is called. The roles associated in these tokens are used for authorization checks. With the decoupling of token generation from \Bindaas, we can add Auth0 login support to any \Bindaas powered service.

\Bindaas can be configured with the configuration file, to use several of its various alternative options for authentication and authorization. A sample \Bindaas configuration file snippet is given below. \textit{api\_key} and \textit{jwt} are the currently implemented authentication protocols of \Bindaas. \textit{OAuthProvider} is the authentication provider class that implements authentication with \textit{jwt}. Similarly, \textit{DBAuthenticationProvider} provides authentication with \textit{api\_key}. The users can also develop their custom authentication provider classes to function with the API Key or JWT based authentication options. The \textit{instanceName} option refers to a specific \Bindaas instance in case of federated or a clustered deployment of \Bindaas. The \textit{host} indicates where the users can access the \Bindaas dashboard and data service APIs from. The default value, ``0.0.0.0'' indicates that the services are accessible from any reachable client. On the other hand, indicating ``127.0.0.1'' or an IP address of the \Bindaas server will limit the incoming requests to only localhost or the networks reachable through the specific network interface.

{\fontsize{8}{8}\selectfont                  
\begin{lstlisting}
{
  "host": "0.0.0.0",  "port": 9099,  "protocol": "http",
  "enableAuthentication": true,
  "enableAuthorization": false,
  "enableAudit": true,
  "authenticationProviderClass": "OAuthProvider",
  "authorizationProviderClass": 
                "AuthorizationProviderImpl",
  "authenticationProtocol": "jwt",
  "auditProviderClass": "DBAuditProvider",
  "proxyUrl": "http://localhost:9099",
  "instanceName": "bindaas"
}
\end{lstlisting}
}

\subsection{Web Console} 

\begin{figure}[ht]
    \begin{center}
            \resizebox{0.75\columnwidth}{!}{
                \includegraphics[width=0.75\textwidth]{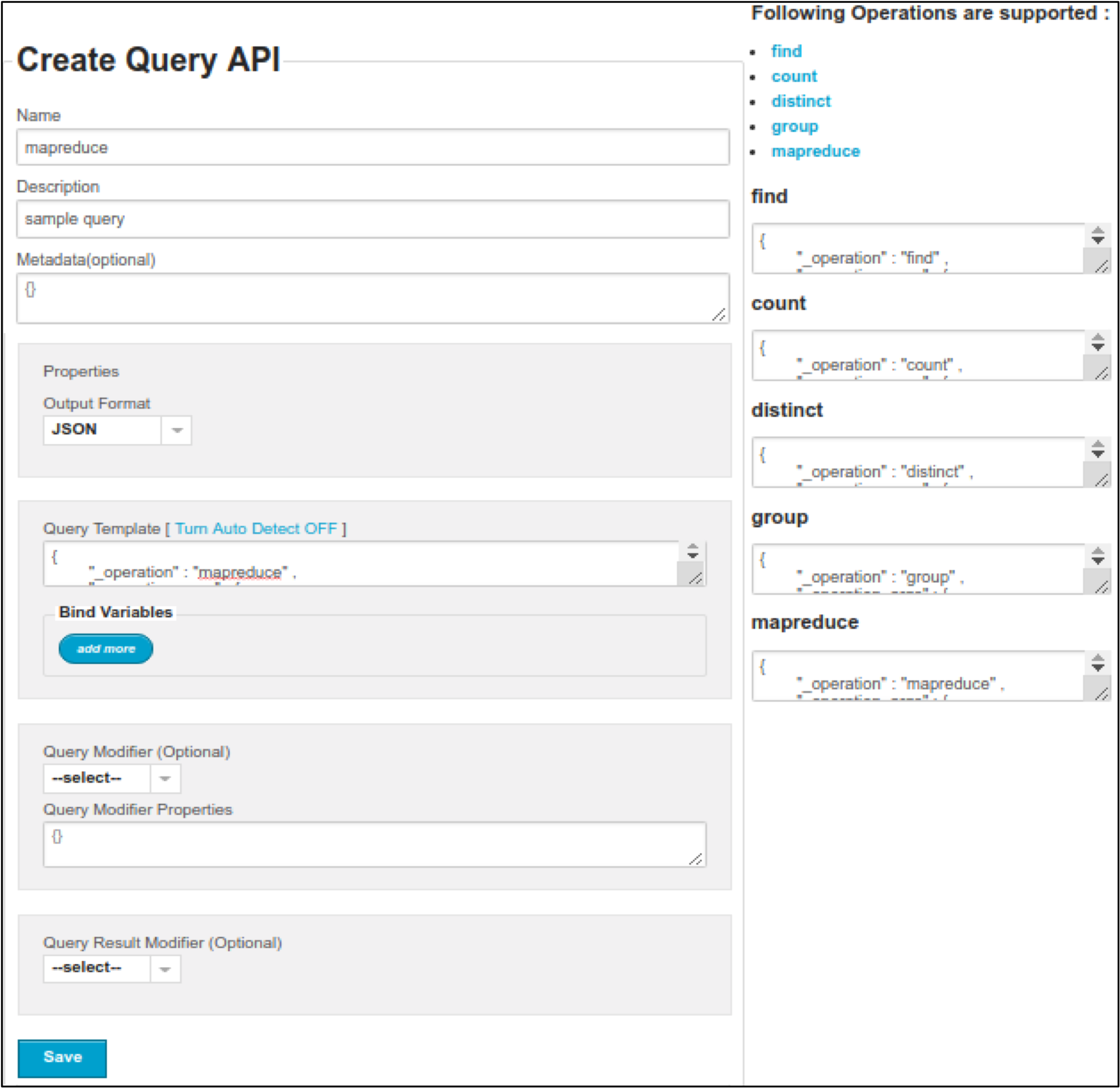}
            }
        \end{center}
        \caption{Create a Query API with the Web Console}
        \label{fig:selectm}
    \end{figure}

\Bindaas web console is developed with Apache Velocity~\cite{gradecki2003mastering} templates for its graphical user interface. \Bindaas uses two different ports: one for the web console to be used typically by the API developers and the administrator. The other one is for the RESTful service interfaces of the data sources. The port of the data services is typically left accessible to the public, whereas the web console port is limited to specific addresses. Figure~\ref{fig:selectm} shows the web console interface to create a query API.

\begin{figure}[ht]
    \begin{center}
            \resizebox{0.75\columnwidth}{!}{
                \includegraphics[width=0.75\textwidth]{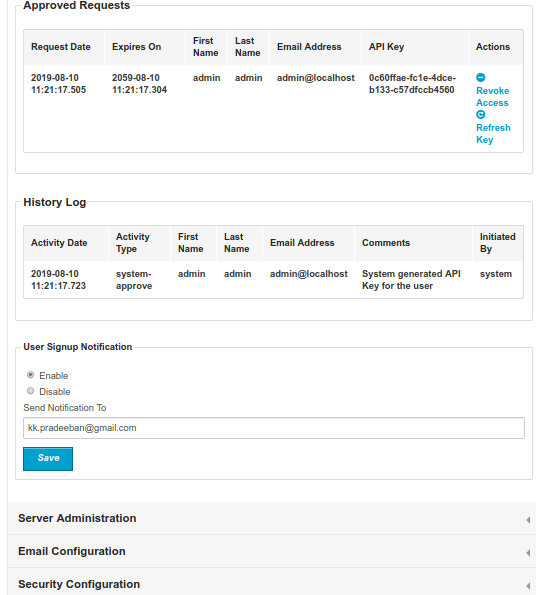}
            }
        \end{center}
        \caption{Administering \Bindaas with its Web Console}
        \label{fig:admins}
    \end{figure}

 Figure~\ref{fig:admins} shows the web console interface for administering \Bindaas, including configuring the middleware framework, activating and deactivating the users, and configuring authentication and authorization of the \Bindaas middleware.

\Bindaas offers an easy to use RESTful interfaces for the big data application developers, without compromising the security aspects provided by the data sources. We also developed toolkits to generate and manage API keys and JWT for \Bindaas, as well as client applications and SDK to consume the \Bindaas data service APIs. Thus, \Bindaas aims to be a complete, extensible, and easy-to-use RESTful web service middleware platform for diverse data sources.
\section{Evaluation}
\label{sec:eval}

We evaluated the \Bindaas middleware for potential overheads that it imposes on its underlying data sources and its scalability to handle a large number of concurrent complex requests. We used an Amazon Web Services (AWS) EC2 memory-optimized cloud VM of instance type m5a.xlarge with 4 vCPUs, 16 GB memory, 8 GB SSD storage, Ubuntu 18.04 operating system, with up to 10 Gbps network interface as our server environment. We used MongoDB 4.0 as our base database server. We then used \Bindaas on top of the database to offer RESTful interfaces to the database server. We populated a MongoDB database with Airbnb data~\cite{airbnb}. We executed database queries on top of the MongoDB directly and then using the \Bindaas data service APIs to evaluate its performance and overhead. We used a computer with 2.8 GHz Intel Core i7 processor, 16 GB 2133 MHz LPDDR3 memory, and 256 GB hard disk, and macOS Sierra operating system as our client environment. We configured Apache JMeter~\cite{halili2008apache} with multiple clients, to emulate several service requests to the Airbnb database in the server, first through the MongoDB interface, and then through the \Bindaas API.

\begin{figure}[ht]
    \begin{center}
            \resizebox{\columnwidth}{!}{
                \includegraphics[width=\textwidth]{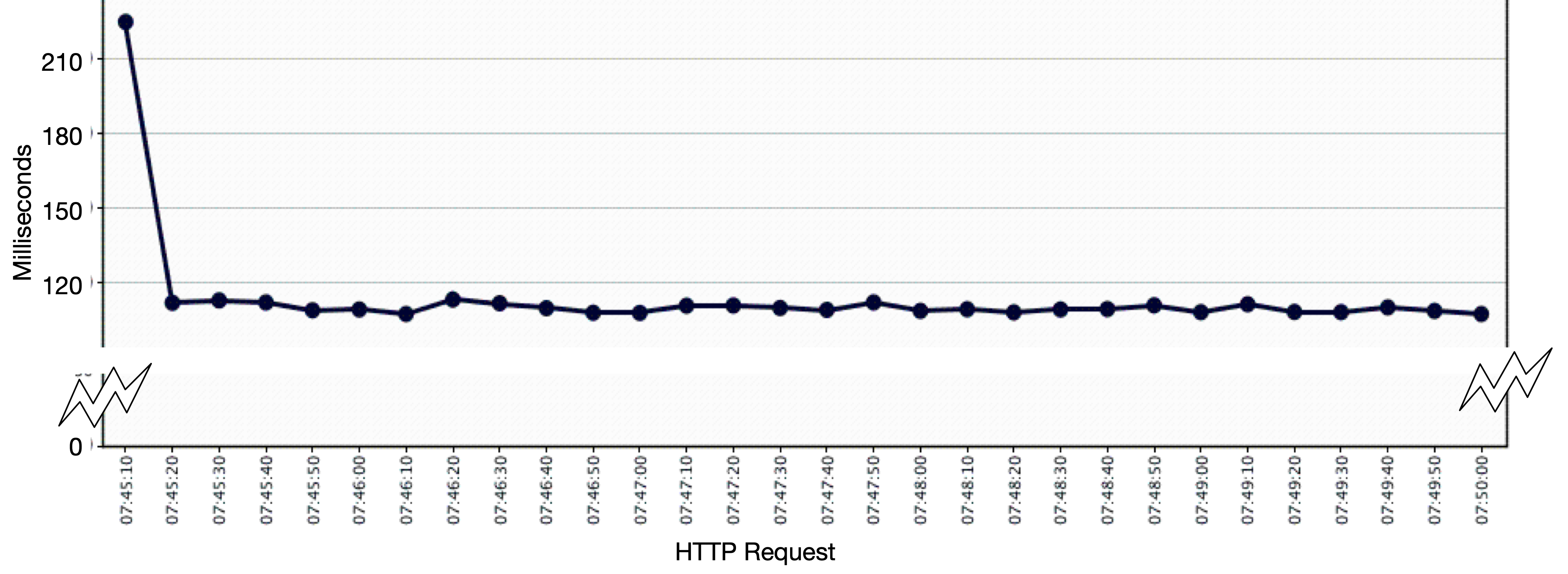}
            }
        \end{center}
        \caption{\Bindaas Response Time of \textit{Count} in a Cloud Server}
        \label{fig:cr}
    \end{figure}

\begin{figure}[ht]
    \begin{center}
            \resizebox{0.85\columnwidth}{!}{
                \includegraphics[width=0.85\textwidth]{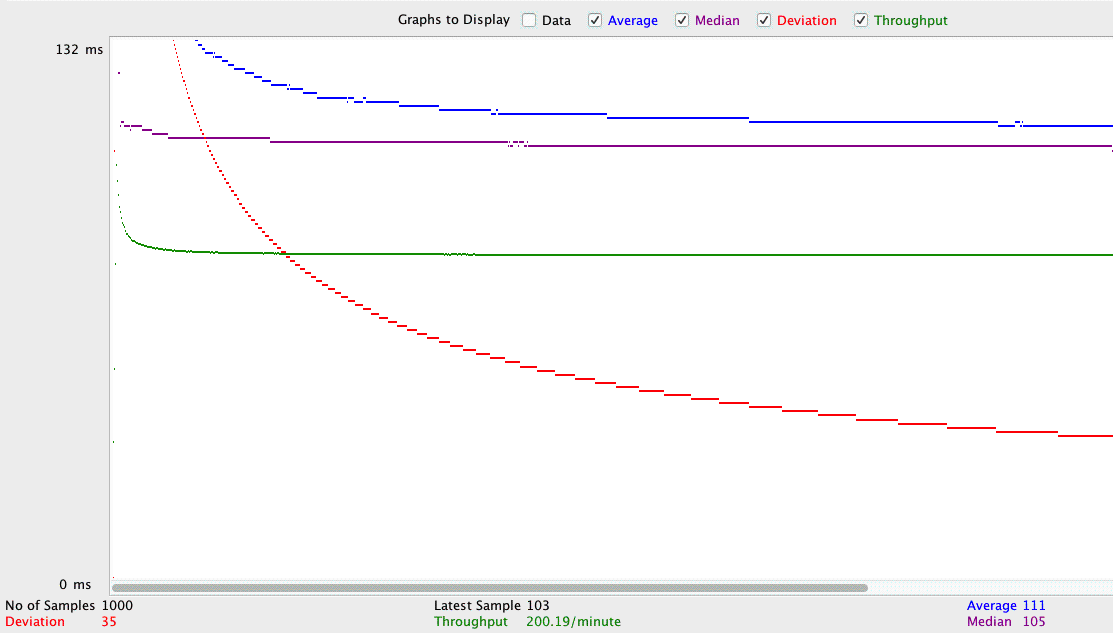}
            }
        \end{center}
        \caption{\Bindaas \textit{Count} Query Performance Statistics in a cloud server}
        \label{fig:cs}
    \end{figure}

First, an additional middleware layer such as \Bindaas in between the data source and the big data application should cause minimal overhead or latency. We observed the efficiency of \Bindaas by comparing the query response time of data sources, directly and then via the \Bindaas RESTful service APIs. We measured the overhead caused by \Bindaas as a difference of query response time of the MongoDB deployment with and without \Bindaas. We observed a marginal difference between the queries executed directly on the Mongo default client and \Bindaas. When compared with a JDBC client for Mongo, the difference was insignificant. We note that this is because \Bindaas itself is merely a JDBC client that internally uses the mongo driver. We also observed similar outcomes with MySQL server as the database.
\begin{figure}[ht]
    \begin{center}
            \resizebox{\columnwidth}{!}{
                \includegraphics[width=\textwidth]{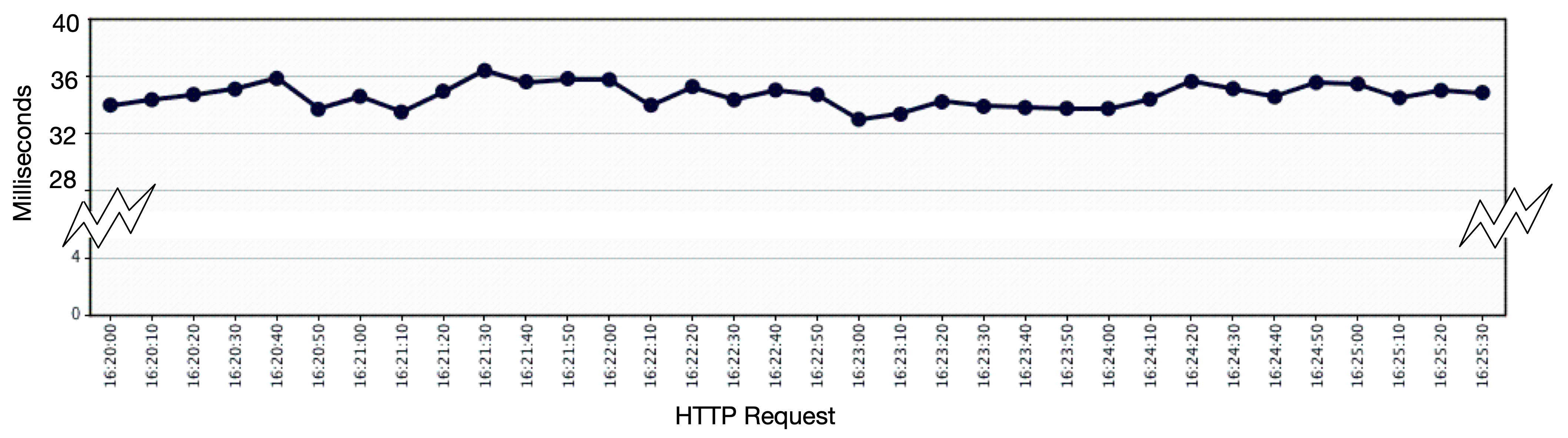}
            }
        \end{center}
        \caption{\Bindaas Response Time of \textit{Count} Locally}
        \label{fig:lr}
    \end{figure}
We first tested our server-client interaction entirely locally in our computer before sending requests to \Bindaas cloud server to evaluate the impact of the network latency in the query execution times. Our query finds the number of hosts who have a profile picture. We assessed the performance of \Bindaas in managing multiple concurrent requests at once. We used Apache JMeter to create several parallel requests at once to the \Bindaas deployment. We emulated 1000 users with 1000 JMeter threads, with a ramp-up period of 300 seconds. Figure~\ref{fig:lr} illustrates the local response time for our query on a MongoDB Airbnb collection of 56 MB, indexed by the host. 
We then repeated the experiment with \Bindaas server with the database environment in the AWS EC2 instance. Figure~\ref{fig:cr} depicts the response time while Figure~\ref{fig:cs} depicts the performance statistics of this request.
    
\begin{figure}[ht]
    \begin{center}
            \resizebox{0.85\columnwidth}{!}{
                \includegraphics[width=0.85\textwidth]{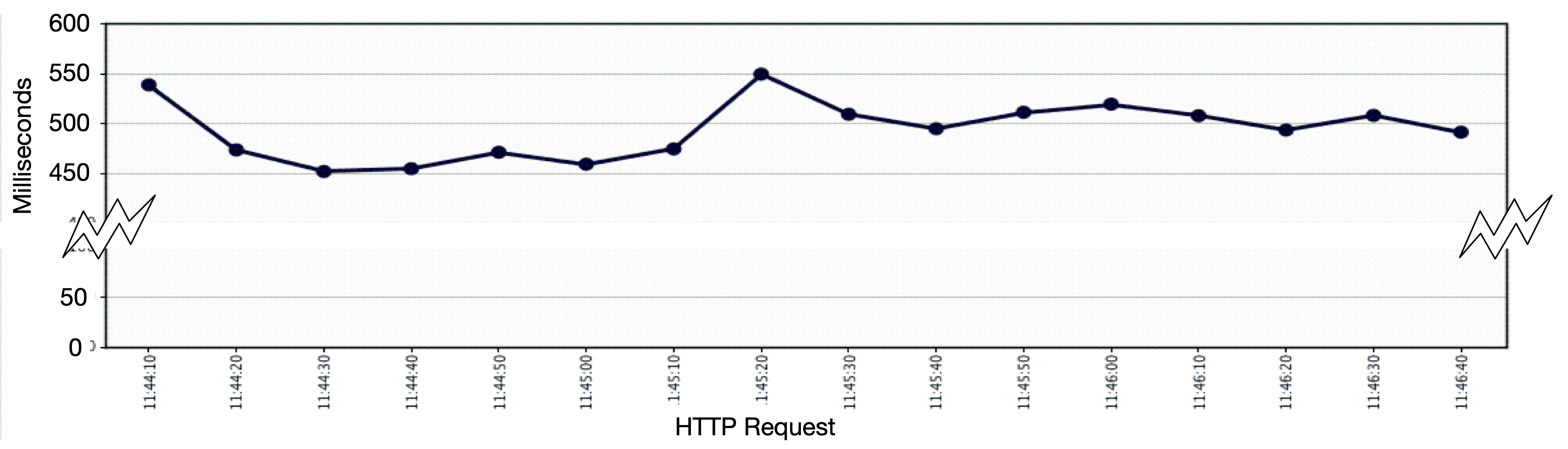}
            }
        \end{center}
        \caption{\Bindaas Response Time of \textit{Distinct}}
        \label{fig:ccr}
    \end{figure}

\begin{figure}[ht]
    \begin{center}
            \resizebox{0.85\columnwidth}{!}{
                \includegraphics[width=0.85\textwidth]{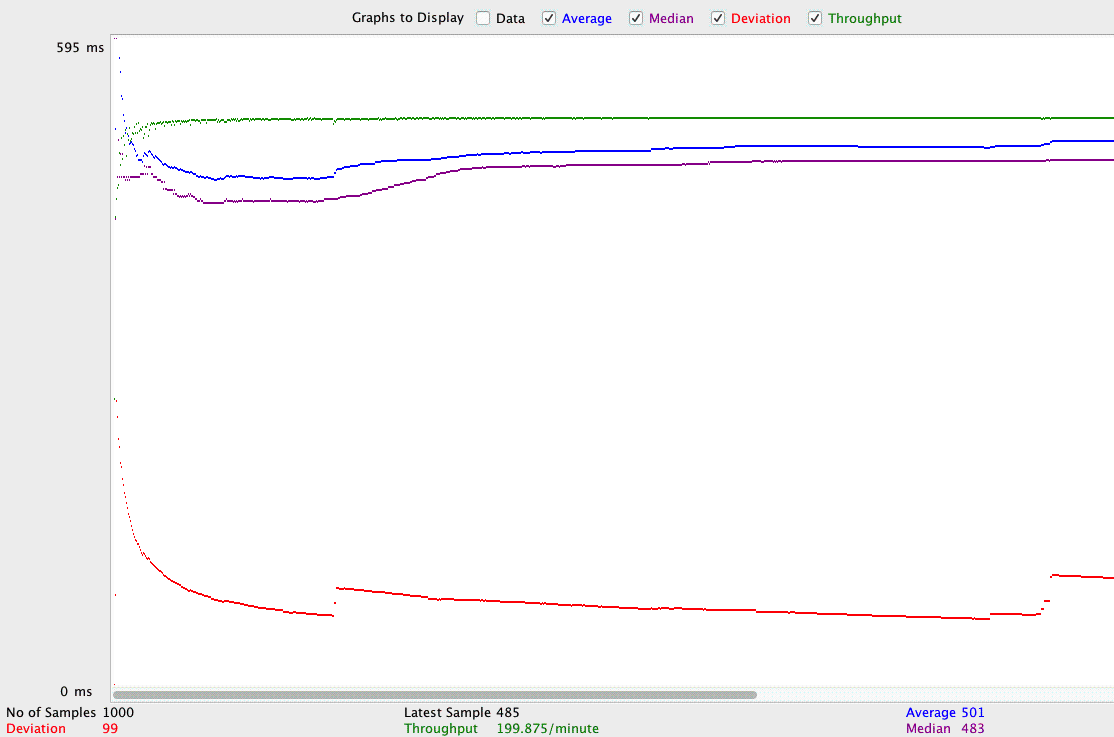}
            }
        \end{center}
        \caption{\Bindaas Query Performance Statistics of \textit{Distinct}}
        \label{fig:ccs}
    \end{figure}

Finally, we evaluated a more complex data service query on \Bindaas deployed on the cloud server, with the same level of concurrency. Our query finds the distinct hosts in the database who satisfy pre-defined criteria. Figure~\ref{fig:ccr} depicts the response time while Figure~\ref{fig:ccs} depicts its performance statistics.\par
\par
Our evaluations highlight that \Bindaas causes minimal latency and can be used to provide RESTful interfaces to the data sources with negligible overhead across multiple queries of different complexities. We observed that \Bindaas effectively handles a large number of concurrent requests without causing a bottleneck or a failure.\par
\par
\section{Conclusion}
\label{sec:concl}
Various application domains use multiple data sources in their big data applications. As data sources consist of diverse interfaces with little to no interoperability among them, these big data applications often end up tied to specific data sources. \Bindaas is a big data middleware framework that offers RESTful interfaces to various relational, NoSQL, and non-traditional data sources. \Bindaas consists of a modular, extensible architecture that lets the developers build big data applications that are not tied to a specific data source. It enables the users to add further features as bundles easily, and also facilitates an easy extension to more data sources, implementing its interfaces. The evaluations demonstrate a minimal overhead when using \Bindaas to create and manage APIs. It has been used in multiple production-grade systems to provide access to various biomedical and clinical resources. 

\Bindaas is under active development with optimizations for more distributed deployments. It is being enhanced to support faster serializations using frameworks such as Protobuf. To support clinical informatics, we are examining ways to support for RDF stores as a new backend storage system. As future work, we aim to build more datasource providers, including cloud storage environments such as AWS S3 buckets, by incorporating cloud clients into \Bindaas. We also propose to deploy \Bindaas in hybrid cloud environments with load balancing, to handle more concurrent loads efficiently. We note that most of these future works can be implemented as bundles and deployed seamlessly into an existing \Bindaas instance.


\footnotesize{\textit{\textbf{Acknowledgements:}} This work was supported by NCI U01 [1U01CA187013-01], Resources for development and validation of Radiomic Analyses and Adaptive Therapy. PI: Ashish Sharma and Fred Prior (Emory, UAMS)
\balance

\bibliography{main}   

%
%

\end{document}